\documentclass[twocolumn,aps]{revtex4-1}
\usepackage{graphicx}
\usepackage{amssymb}
\usepackage{amsfonts}
\usepackage{amsmath}
\usepackage{amsbsy}
\usepackage{natbib}
\usepackage{comment}

\usepackage{color}
\usepackage{float}

\renewcommand{\vec}[1]{\mbox{\boldmath$\mathrm{#1}$}}
\newcommand{\be}{\begin{equation}}
\newcommand{\ee}{\end{equation}}
\newcommand{\ben}{\begin{eqnarray}}
\newcommand{\een}{\end{eqnarray}}

\begin{document}

\title{Influence of spin-orbit and spin-Hall effects on the  spin Seebeck current beyond  linear response: A Fokker-Planck approach}

\author{L. Chotorlishvili$^1$, Z. Toklikishvili$^2$, X.-G. Wang$^{3}$, V.K. Dugaev$^{4}$, J. Barna\'{s}$^{5,6}$ J. Berakdar$^1$}

\address{$^1$ Institut f\"ur Physik, Martin-Luther Universit\"at Halle-Wittenberg, D-06120 Halle/Saale, Germany \\
$^2$  Faculty of Exact and Natural Sciences, Tbilisi State University, Chavchavadze av.3, 0128 Tbilisi, Georgia\\
$^3$ School of Physics and Electronics, Central South University, Changsha 410083, China\\
$^{4}$ Department of Physics and Medical Engineering, Rzeszow University of Technology, 35-959 Rzeszow, Poland\\
$^{5}$Faculty of Physics, Adam Mickiewicz University, ul. Umultowska 85, 61-614 Poznan, Poland\\
$^{6}$Institute of Molecular Physics, Polish Academy of Sciences, ul. M. Smoluchowskiego 17, 60-179 Pozna\'{n}, Poland}

\begin{abstract}
We study the spin transport theoretically in heterostructures consisting of a ferromagnetic metallic thin film sandwiched between heavy-metal and oxide layers.  The spin current in the heavy metal layer is generated via the spin Hall effect, while the oxide layer induces at the interface with the ferromagnetic layer a spin-orbital coupling of the Rashba type. Impact of the spin Hall effect and Rashba spin-orbit coupling on the spin Seebeck current is explored with a particular emphasis on nonlinear effects. Technically, we employ the Fokker-Planck approach and contrast the analytical expressions with full numerical micromagnetic simulations. We show that when an external magnetic field $H_{0}$  is aligned parallel (antiparallel) to the Rashba field, the spin-orbit coupling enhances (reduces) the spin pumping current. In turn, the spin Hall effect and the Dzyaloshinskii-Moriya interaction are shown to increase the spin pumping current.
\end{abstract}
\date{\today}
\maketitle
\section{Introduction}

In a seminal paper \cite{Bychkov}, Bychkov and Rashba explored the impact of spin-orbit (SO) interaction on the properties of two-dimensional semiconductor heterostructures. Since then, the basic idea of Bychkov and Rashba was carried over to other research areas of physics. It was shown, for instance, that the SO interaction plays a significant role in the quantum spin Hall Effect in graphene \cite{Kane}, Bose-Einstein condensates \cite{Spielman}, and in the orbital-based electron-spin control \cite{Efros}. Recent experiments \cite{Miron, Haazen} revealed the role of SO interaction in the motion of domain walls, as well. Combining the SO coupling and thermal effects bring in new insight and phenomena. A thermal bias applied to a ferromagnetic insulator leads to the formation of a thermally assisted magnonic spin current that is proportional to the temperature gradient. This phenomenon falls in the class of spin Seebeck effects and may be useful for  thermal control of  magnetic moments \cite{Kovalev,Ritzmann,Bauer1,Barker,Saitoh1,Kehlberger,Schreier,Boona,Basso,Lefkidis,Rezende,Brechet,Bartell,Ansermet,sayyed,chotor,xguang}.
The objective of this paper is to study the impact of SO interaction on the formation and transport of thermally assisted magnonic spin current in spin-active multilayers.  We investigate two different heterostructures which include a layer of ferromagnetic metal sandwiched between heavy metal and oxide materials, see Fig. \ref{fig1} and Fig. \ref{fig2}. In both cases, an inversion asymmetry is caused by two different interfaces -- heavy-metal/ferromagnet and ferromagnet/oxide ones. A large SO coupling is present in the heavy metal \cite{Wang, Nagaosa, Buhrman, Emori, Kim}. This study is motivated by the experimental work in Ref. \cite{Emori} with a particular  attention to the systems Pt/Co/AlO$_{x}$ and Ta/CoFeB/MgO.
Moreover, the torques generated by strong SO coupling are generally different from the Slonczewski's spin-transfer torque \cite{Wang, Lavrijsen, Rashba}, with the prospect for novel physical effects in the heavy-metal/ferromagnetic-metal/oxide heterostructures. An applied an electric voltage (see Fig.1) generates charge current in the ferromagnetic and heavy metal layers. This current in the heavy-metal layer leads to spin current due to the spin Hall effect, which is then injected into the thin ferromagnetic layer \cite{Gambardella, Dyakonov, Hirsch, Sugai, Liu, Pai, Kondou, Seo} and acts as an extra torque on the localized magnetic moments in the ferromagnet. The induced torque influences the magnetization dynamics, which is the topic of this work. To describe the influence of the spin current on the magnetization dynamics in the ferromagnetic layer we add a relevant term to the Landau-Lifshitz-Gilbert (LLG) equation.
%The spin Hall effect (SHE) arises due to the proximity with heavy metal, while
In turn, the Rashba SO coupling at the ferromagnet/oxide interface in the presence of the charge current results in   a spin polarization at the interface, with the exchange coupling exerting  a torque on the ferromagnetic layer, as well. Thus, the SHE and the Rashba SO coupling influence the magnetization dynamics in the ferromagnetic layer through the Rashba and SHE torques, both incorporated into the LLG equation (the Rashba and SHE fields).
The considered setup allows to formally investigate the interplay/competition of the torques due to Rashba  SO interaction and SH effect. The Rashba SO torque  acts  field-like, while the torque due to the spin current generated {\it via} the spin Hall effect is predominantly of  damping/antidumping like in nature.
We utilize the Fokker-Planck method \cite{Z.LiandS.Zhang} for the stochastic LLG equation for studying the magnetic dynamics beyond the linear response regime.
The influence of the  Rashba-type SO coupling on the magnonic spin current was studied in the works \cite{Okuma and K. Nomura,Farzad Mahfouzi,Branislav K. Nikolic}.

In the system shown in Fig.1, the normal metal with temperature $T_{N}$ is attached to the ferromagnet with  $T_{F} > T_{N}$. We consider the spin current flowing from the ferromagnetic to a normal metal layer.
Magnons from the high-temperature region diffuse to the lower temperature part giving rise to a magnonic spin current
and thus also to the spin Seebeck effect (SSE) \cite{Saitoh, Uchida,Xiao}. Magnonic spin current pumped from the ferromagnet into the
normal metal, $I_{sp}$, increases with the temperature difference, $I_{sp}\sim T_{F}-T_{N}$.
However, the spin current injected from the ferromagnetic layer to the normal metal is not the only spin current that crosses the normal-metal/ferromagnet interface.
The fluctuating spin current $I_{fl}$ is generated in the normal metal and flows towards the ferromagnet, i.e. in the direction opposite to the magnonic spin pumping current.
The quantity of interest is therefore the total spin current, $I_{tot}=I_{sp}+I_{fl}$ that crosses the normal-metal/ferromagnet interface.
We show that   $I_{tot}$ is drastically influenced by the proximity of the heavy metal (due to spin Hall effect) and the oxide  (due to Rashba spin-orbit coupling).
In the second system (see Fig.2) an additional normal-metal layer is attached to the ferromagnetic one.

%spin- lack of the inversion symmetry that arises because of the ferromagnet/ oxide interface.
%
%The magnetization dynamics in the ferromagnet is described by stochastic LLG equation with additional spin torque terms as already described above. The spin torque terms influence magnetization dynamics in the ferromagnet and therefore change the spin currents. We explore the impact of the SO coupling and SHE on the spin Seebeck (SSE) current.
%The spin current emitted from the ferromagnet to the normal metal is evaluated in the framework of the Fokker-Plank equation approach for the stochastic LLG equation.\cite{Chtorlishvili}.
%Effect of the DMI term is explored numerically in micromagnetic simulations.

The paper is organized as follows. In Sec. \textbf{II} we introduce the model under consideration. In Sec. \textbf{III} and \textbf{IV} we explore the spin current in two different heterostructures.  For the sake of simplicity, we neglect Dzyaloshinskii-Moriya interaction (DMI).
Effects of the DMI term and magnetocrystalline anisotropy are explored numerically by micromagnetic simulations and are described
in Sec.  \textbf{V}.
%and \textbf{VI}
%we study the effects of DM interaction
%and magnetocrystalline anisotropy respectively,
Section \textbf{VI} summarizes the findings. Main technical details are deferred to the appendices.

\section{Theoretical model}

For the heavy-metal/ferromagnetic-metal/oxide sandwich we choose the ferromagnetic metallic layer to be in direct contact with a nonmagnetic metallic layer, as shown in Fig.1. We also assume that, due to a strong electron-phonon interaction, the local thermal equilibrium between electrons and phonons in both ferromagnetic and normal-metal layers is established,   $T_{F}^{p}=T_{F}^{e}=T_F$ and $T_{N}^{p}=T_{N}^{e}=T_{N}$. The magnon temperature $T_{F}^{m}$ in the ferromagnetic layer  differs  in general from the  temperature of electrons/phonons, $T_{F}^{m}\neq T_{F}$ \cite{Xiao}.

\begin{center}
   \begin{figure}[!t]
    \centering
    \includegraphics[width=0.48\textwidth]{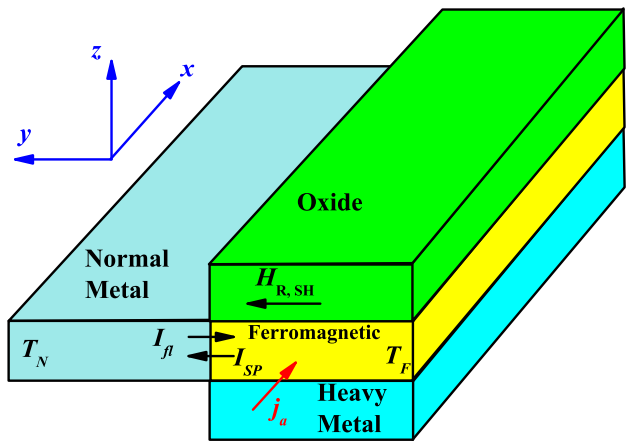}
        \caption{Schematic illustration of the system. A ferromagnetic metallic layer is sandwiched between the oxide and heavy metal layers. The injected current $j_{a}$ flows in the ferromagnetic and heavy metal layers in the $x$ direction.
        The Rashba field $H_{R}$ and the spin Hall field are oriented along the  $y$ axis. The normal metal with the temperature $T_{N}$ is attached to the ferromagnetic layer. The temperature
        of the ferromagnetic layer $T_{F}$ is different from  $T_{N}$. }
    \label{fig1}
    \end{figure}
\end{center}

At nonzero temperatures, the thermally activated magnetization dynamics in the ferromagnet gives rise to a spin current flowing into the normal metal. This effect is known as spin pumping~\cite{Saitoh, Foros, Tserkovniak, Adachi}. The corresponding expression for the spin current density reads \cite{Xiao,1Chtorlishvili}
\begin{equation}\label{spinpump}
\vec{I}_{sp}(t)=\frac{\hbar}{4\pi}[g_{r}\vec{m}(t)\times \dot{\vec{m}}(t)+g_{i}\dot{\vec{m}}(t)],
\end{equation}
where $g_{r}$ and $g_{i}$ are the real and imaginary parts of the dimensionless spin mixing conductance of the ferromagnet/normal-metal $(F|N)$ interface, while $\vec{m}(t)=\vec{M}(t)/M_s$ is the dimensionless unit vector along the magnetization orientation (here $M_{s}$ is the saturation magnetization) and $\dot{\vec{m}}\equiv d\vec{m}/dt$.
The spin current is a tensor describing the spatial distribution of the current flow and orientation of the flowing spin (magnetic moment).
Due to the geometry of the system under consideration, the spin current flows along the $y$-axis, see Fig. \ref{fig1}.
In turn,  the spin polarization of the current depends on the orientation of the magnetic moment and its time derivative.  The average spin depends on the ground state magnetic order which in our case is collinear with the external magnetic field (applied along the $y$-axis).
Therefore, the only nonzero component of the average spin current tensor is ${I}_{sp}^{y}$.

Thermal noise in the normal-metal layer activates a fluctuating spin current flowing from the normal metal to the ferromagnet \cite{Foros},
\begin{equation}\label{NouseSpinCurrent}
\vec{I}_{fl}(t)=-\frac{M_{s}V}{\gamma}\vec{m}(t)\times\vec{\zeta}'(t).
\end{equation}
Here, $V$ is the total volume of the ferromagnet, $\gamma$ is the gyromagnetic factor, and
$\vec{\zeta}'(t)=\gamma\vec{h}'(t)$ with $\vec{h}'(t)$ denoting the random magnetic field. In the classical limit, $k_{B}T\gg\hbar\omega_{0}$, the correlation function $\langle \zeta'_{i}(t)\zeta'_{j}(t')\rangle$  of
$\vec{\zeta}'(t)$ reads
\begin{equation}\label{correlation1}
\langle \zeta'_{i}(t)\zeta'_{j}(t')\rangle =\frac{2\alpha'\gamma k_{B}T_{N}}{M_{s}V}\delta_{ij}\delta(t)\equiv \sigma'^2\delta_{ij}\delta(t),
\end{equation}
where $\langle...\rangle$ denotes the ensemble average, and $i,j=x,y,z$.
Furthermore,  $\omega_{0}$ is the ferromagnetic resonance frequency and $\alpha'$ is the contribution to the damping constant due to  spin pumping, $\alpha^\prime =\gamma\hbar g_{r}/4\pi M_{s}V$.  We emphasize that the correlator  (Eq.(\ref{correlation1})) is proportional to the temperature $T_{N}$.

The total spin current flowing through the ferromagnet/normal-metal interface is given by the sum of pumping and fluctuating spin currents, $\vec{I}_{tot}=\vec{I}_{sp}+\vec{I}_{fl}$. For clarity of notation, we omit here (and also in the following) the time dependence of spin currents, normalized magnetization, random magnetic fields, and their correlators. This dependence will be restored if necessary.
According to Eqs. (\ref{spinpump}) and (\ref{NouseSpinCurrent}), the total average spin current flowing across the interface can be written in the following form \cite{Xiao}:
\begin{equation}\label{totalspincurrent}
\langle\vec{I}_{tot}\rangle=\frac{M_{s}V}{\gamma}[\alpha'\langle\vec{m}\times\dot{\vec{m}}\rangle-\langle\vec{m}\times\vec{\zeta'}\rangle].
\end{equation}

Now, we assume that a spatially uniform current of density $\vec{j}_{a}=j_{a}\vec{i}_{x}$ is injected along the $x$-axis. This current  gives rise to additional torques owing to the spin Hall effect and Rashba spin-orbit interaction. Thus, the  magnetization dynamics is then modified and is governed by the stochastic LLG equation \cite{Emori}:
\begin{equation}\label{LLG}
\frac{d\vec{m}}{dt}=-\gamma\vec{m}\times(\vec{H}_{\rm eff}+\vec{h})+\alpha\vec{m}\times\dot{\vec{m}} + \vec{\tau}_{SO},
\end{equation}
%where $\vec{m}(\vec{r},t)=\vec{M}(\vec{r},t)/M_{s}$ is the normalized local magnetization {\color{blue} Comment: below Eq.1 \bf m is defined as \bf m/M}, %$\gamma$ is the gyromagnetic ratio,
where $\alpha$ is the Gilbert damping constant, $\vec{h}$ is the time-dependent random magnetic field in the ferromagnet, and $\vec{H}_{\rm eff}$ is an effective field.
This effective field consists of three contributions: the exchange field, the external magnetic field oriented along the $y$-axis, and the field corresponding to the DM interaction:
\begin{eqnarray}
\label{effectivefield}
&&\vec{H}_{\rm eff}=\frac{2A}{\mu_{0}M_{s}}\nabla^{2}\vec{m}+H_{0}\vec{y}-\frac{1}{\mu_{0}M_{s}}\frac{\delta E_{DM}}{\delta \vec{m}},\nonumber\\
&&E_{DM}=D\big[m_{z}\vec{\nabla}\vec{m}-\big(\vec{m}\vec{\nabla}\big)m_{z}\big].
\end{eqnarray}
For the sake of simplicity, in the analytical part we take into account  only the external magnetic field.
In turn, the term $\vec{\tau}_{SO}$ in Eq.(5)  describes SO torques related to the Rashba SO coupling and the spin Hall effect,
\begin{eqnarray}\label{SOT}
\vec{\tau}_{SO}=&&-\gamma\vec{m}\times\vec{H}_{R}+\gamma\eta\xi\vec{m}\times(\vec{m}\times\vec{H}_{R})
\\&&+\gamma\vec{m}\times(\vec{m}\times \vec{H}_{SH}),\nonumber
\end{eqnarray}
where $\xi$ is a non-adiabatic parameter, and $\eta=1$ when the torque has Slonczewski-like form, while $\eta=0$ in the opposite case \cite{Emori}.
In the above equation, the DM interaction enters the effective magnetic field, while the effect of
Rashba SO coupling and spin Hall effect are included by means of the extra torque added to the LLG equation.

As already mentioned above, the charge current flowing in the thin ferromagnetic layer leads to spin polarization at the ferromagnet/oxide interface. The accumulated spin density in vicinity of the interface interacts with the local magnetization by means of the exchange coupling. This effect may be described by an effective Rashba field $\vec{H}_{R}=H_{R}\vec{i}_{y}$ \cite{Wang,Miron,Gambardella}:
\begin{equation}\label{Rashba}
\vec{H}_{R}=\frac{\alpha_{R}P}{\mu_{0}\mu_{B}M_{s}}(\vec{i}_{z}\times\vec{j}_{a})=\frac{\alpha_{R}Pj_{a}}{\mu_{0}\mu_{B}M_{s}}\vec{i}_{y},
\end{equation}
where $\alpha_{R}$ is the Rashba parameter  and $P$ is the degree of spin polarization of conduction electrons \cite{Gambardella}.
The first term in  Eq.(\ref{SOT}) corresponds to the out-of-plane torque and is related to the effective field $H_{R}$.  This torque  is oriented perpendicularly to the $(\vec{m},\vec{H}_{R})$ plane. The second term in Eq.(\ref{SOT})  captures the effects of spin diffusion inside the magnetic layer and the spin current associated with the Rashba interaction at the interface. For more details, we refer to the work \cite{Wang}.

The last term in Eq.(\ref{SOT}) corresponds to the spin Hall torque \cite{Dyakonov, Hirsch}, expressed by the spin Hall field $\vec{H}_{SH}$.
The spin current is generated due to the spin Hall effect in the heavy metal layer and is injected into the ferromagnetic layer.  For more details, we refer to the references \cite{Liu, Pai, Kondou, Seo}.
The explicit expression for $\vec{H}_{SH}$ reads:
\begin{equation}\label{SH}
\vec{H}_{SH}=\frac{\hbar \theta_{SH}j_{a}}{\mu_{0}2eM_{s}L_{z}}\vec{i}_{y},
\end{equation}
where $L_{z}$ is the thickness of the ferromagnetic layer, while $\theta_{SH}$ is the spin Hall angle (defined as the ratio of spin current and charge current densities).

As already mentioned above, total random magnetic field $\vec{h}(t)$ has two contributions from  different noise sources: the thermal random field $\vec{h}_{0}(t)$, %{\color{red} the damping constant is $\alpha_{0}$,}
and the random field $\vec{h}'(t)$.
% of the normal metal.
%{\color{red}, damping constant $\alpha'$}.
Since the random fields are statistically independent, their correlators are additive and fully determined by the total (enhanced) magnetic damping $\alpha=\alpha_{0}+\alpha'$ \cite{Xiao} (with $\alpha_0$ being the damping parameter of the ferromagnetic material, i.e., without contributions from pumping currents),
\begin{equation}\label{correlation2}
\langle\zeta_{i}(t)\zeta_{j}(0)\rangle=\frac{2\alpha\gamma k_{B}T_{F}^{m}}{M_{s}V}\delta_{ij}\delta(t)=\sigma^{2}\delta_{ij}\delta(t),
\end{equation}
where $\vec{\zeta}(t)=\gamma\vec{h}(t)$, and $\alpha T_{F}^{m}=\alpha_{0}T_{F}+\alpha' T_{N}$.
%{\color{blue} Comment: Check definitions in the manuscript  of $\alpha$, $\alpha_0$ and $\alpha^\prime$.}

\section{Spin Current:  N/F structure}

%In Fig.\ref{fig1} is shown a schematic Illustration of the system oxide/ferromagnetic/heavy metal sandwich. Two interfaces oxide/ferromagnetic and ferromagnetic/heave metal leads to the structural asymmetry and interface with heavy metal contributes to the SO coupling. The injected current $j_{a}$ flows in the $x$ direction.
%The SO coupling produces two main effects: the Rashba effective field \cite{Gambardella} and spin Hall effect.

The injected electrical current creates a transverse spin current in the heavy-metal layer via the spin Hall effect (or spin accumulation at the boundaries of the sample) \cite{Saitoh}.  In turn, the Rashba SO interaction in the presence of charge current gives rise to additional torque as already described above. In the case under consideration, the Rashba SO field, Eq.(\ref{Rashba}), and the spin Hall field,   Eq.(\ref{SH}),  are oriented along the $y$ axis.    When temperature of the ferromagnetic film differs from that of the normal metal, $T_{F}\neq T_{N}$, the spin Seebeck current emerges in the Fe/N contact. Note, this current also exists in the absence of spin-orbit interaction and for $j_a=0$. Below, we calculate the total spin current in the N/F structure, taking into account the Rashba  SO field and the spin Hall effect.

In order to calculate the spin pumping current, $\langle\vec{I}_{sp}\rangle=\frac{M_{s}V}{\gamma}\alpha'\langle\vec{m}\times\dot{\vec{m}}\rangle$, we use Eq.(\ref{LLG1}) (see Appendix A) and find
\begin{equation}\label{Spincurrent}
\langle\vec{I}_{sp}\rangle=\alpha'\frac{M_sV}{\gamma}(-\langle\vec{m}\times\vec{\omega}_2\rangle-\langle\vec{m}\times\vec{m}\times\vec{\omega}_1\rangle) ,
 \end{equation}
where $\vec{\omega}_1$ and $\vec{\omega}_2$ are defined in the Appendix A, see Eq.(\ref{LLG1}).
Utilizing Eq.(\ref{Omega}) and Eq.(\ref{Stationarysolution}) we find mean values of the magnetization components (see Appendix B):
\begin{eqnarray}\label{mean values}
\langle m_y\rangle=-L(\beta\omega_2),~~~\langle m_{y}^2\rangle=1-\frac{2L(\beta\omega_2)}{\beta\omega_2},\nonumber \\
\langle m_{x}^2\rangle=\langle m_{z}^2\rangle=\frac{L(\beta\omega_2)}{\beta\omega_2},
\end{eqnarray}
where $\beta=2/\sigma^2$, and $L(x)=\coth x-\frac{1}{x}$ is the Langevin function.
From Eq.(\ref{mean values}) we obtain $\langle\vec{m}\times\vec{\omega}_2\rangle=0$, and $\langle(\vec{m}\times\dot{\vec{m}})_x\rangle=0$,
$\langle(\vec{m}\times\dot{\vec{m}})_y\rangle=-\omega_1(1-\langle m_y^2)\rangle)$, $\langle(\vec{m}\times\dot{\vec{m}})_z\rangle=0$.
Thus, the only nonzero component of the spin pumping current is $\vec{I}_{sp}^{y}$,
\begin{eqnarray}\label{Spincurrenty}
\langle\vec{I}_{sp}^{y}\rangle=&&\alpha'\frac{M_sV}{\gamma} \omega_1(1-\langle m_y^2\rangle)\nonumber \\
&&=\alpha'\frac{M_sV}{\gamma}\frac{2\omega_1}{\beta\omega_2}L(\beta\omega_2).
\end{eqnarray}

For the evaluation of the fluctuating spin current $\langle\vec{I_{fl}}\rangle=-\frac{M_sV}{\gamma}\langle \vec{m}\times\zeta'\rangle$, we linearize the LLG equation, Eq.(\ref{LLG1}), near to the equilibrium point: $\langle m_x\rangle=\langle m_z\rangle=0$, $\langle m_y\rangle=-L(\beta\omega_2)$:
\begin{eqnarray}\label{LinearezedLLG}
&&\dot{m}_x=\omega_1 m_z +\omega_2 \langle m_y\rangle m_x-\langle m_y\rangle \zeta_z(t), \nonumber \\
&&\dot{m}_z=-\omega_1 m_z +\omega_2 \langle m_y\rangle m_z+\langle m_y\rangle \zeta_x(t).
\end{eqnarray}
Fourier transforming to the frequency domain ($\tilde{g}=\int ge^{i \omega t}dt$ and $g=\int \tilde{g}e^{-i\omega t}d\omega/2\pi$),  from  Eq.(\ref{LinearezedLLG}) we obtain $\tilde{m}_i(\omega)=\sum_{j}\chi_{ij}(\omega)\tilde{\zeta}_j(\omega)$, where $i,j=x,z$, and
\begin{eqnarray}
\chi_{ij}(\omega) =\frac{\langle m_y\rangle}{(\omega_2\langle m_y\rangle+i\omega)^2+\omega_1^2} \nonumber\\
\times\begin{pmatrix} \omega_1 & (\omega_2\langle m_y\rangle+i\omega)\\ -(\omega_2\langle m_y\rangle+i\omega) & \omega_1  \end{pmatrix},
\end{eqnarray}
\begin{equation}\label{fluctuationtransform}
\langle m_i(t)\zeta'_x(0)\rangle=\sigma'^2\int^{+\infty}_{-\infty}\chi_{ij}(\omega)e^{-i\omega t}\frac{d \omega}{2 \pi} .
\end{equation}
Equation(\ref{fluctuationtransform}) has nonzero elements:
\begin{eqnarray}\label{nonzerotransform}
&&\langle m_z(t)\zeta'_{x}(0)\rangle=-\langle m_x(t)\zeta'_z(0)\rangle \nonumber\\
&&=-\sigma'^2\langle m_y\rangle\int_{\infty}^{+\infty}\frac{\omega_2\langle m_y\rangle+i\omega}{(\omega_2\langle m_y\rangle+i\omega)^2+\omega_1^2}e^{-i\omega t}\frac{d \omega}{2\pi} .
\end{eqnarray}
Details of  calculating the  integral in Eq.(\ref{nonzerotransform}) are presented in Appendix C. Taking into account Eq.(\ref{nonzerotransform}) one obtains
\begin{equation}\label{Corelator}
\langle\vec{m}\times\vec{\zeta}'\rangle_y=\langle m_z\zeta'_x-m_x\zeta'_z\rangle=-\sigma'^2L(\beta\omega_2).
\end{equation}
The fluctuating spin current has only one nonzero component, i.e., the $y$ component -- similarly as the spin pumping current does,
\begin{equation}\label{fluctuatingcurrent}
\langle\vec{I}_{fl}^{y}\rangle=\frac{M_{s}V}{\gamma}\sigma'^{2}L(\beta\omega_2).
\end{equation}
We emphasize  that when calculating the spin pumping current, we did not employ  a  linearization procedure. Accordingly, the expression for the spin pumping current, Eq.(\ref{Spincurrenty}),
is valid for an arbitrary deviation of the magnetization from the ground state magnetic order, even for thermally assisted magnetization-reversal instability processes, meaning the
transversal components $m_{x},~m_{y}$ can be arbitrarily large. On the other hand, the expression for the fluctuation spin current, Eq.(\ref{fluctuatingcurrent}), was obtained upon a linearization near the equilibrium point, as described at the beginning of this paragraph.
Taking into account the above derived formula %Eq.(\ref{totalspincurrent}) and
Eq.(\ref{Spincurrenty}) and Eq.(\ref{fluctuatingcurrent}) for spin pumping and fluctuation currents, respectively, we deduce the following expression of the total spin current:
\begin{equation}\label{totalspincurrenty}
\langle\vec{I}_{tot}^{y}\rangle=\frac{M_sV}{\gamma}L(\beta\omega_2)\big[\alpha'\frac{\sigma^2 \omega_1}{\omega_2}+\sigma'^2\big].
\end{equation}
When  $\vec{H}_{\rm eff}=(0,H_0,0)$, where $H_0$ is the external magnetic field oriented along the $y$ axis \cite{Miron}, then using Eq.(\ref{correlation1}), Eq.(\ref{correlation2}) and Eq.(\ref{Omega}) one obtains from Eq.(\ref{totalspincurrenty}),  \begin{eqnarray}\label{totalspincurrentyExplicit}
&&\langle I_{tot}^{y}\rangle=2\alpha' k_{B}L\bigg(\frac{M_{s}V(\alpha H_{0}+(\alpha-\eta\xi) H_R-H_{SH})}{\alpha k_B T_{F}^{m}}\bigg) \nonumber\\
&&\times\bigg(\frac{\alpha(H_{0}+H_{R}+\alpha H_{SH})T_{F}^{m}}{\alpha H_{0}+(\alpha -\eta\xi) H_{R}-H_{SH}}-T_{N}\bigg),
\end{eqnarray}
where $\eta=0,1$ and we inspect in the following the $\eta=0$ case.

We analyze now in more details Eq.(\ref{totalspincurrentyExplicit}) for $\eta =0$ and for several asymptotic cases. Let us begin with the case of a negligible spin Hall effect.
Assuming a small $H_{SH}$, $H_{SH}\ll\alpha H_{R},~\alpha H_{0}$, we derive from Eq.(\ref{totalspincurrentyExplicit}) the spin current
in the following two regimes: (i) The low temperature regime, $M_{s}V(H_{0}+H_R)/{k_B T_{F}^{m}}\gg1$, and (ii) the high temperature regime, $M_{s}V(H_{0}+H_R)/k_B T_{F}^{m}\ll1$. These two regimes can be  equivalently referred to as the high and weak magnetic field limits, respectively.
%The spin-orbit torque exerted on magnetic moment is then of field-like character.
In particular, in the low temperature limit, upon
taking into account the property of the Langevin function, $L\big(x\big)=\coth\big(x\big)-1/x$, $L\big(x\gg1\big)\approx1$,
we find that the spin current depends neither on the SO coupling nor on the external magnetic field, $\langle I_{tot}^{y}\rangle=2\alpha' k_{B}(T_{F}^{m}-T_{N})$, and is solely determined by the temperature bias. In the low-temperature regime (strong magnetic field), the magnetic fluctuations are small and the spin current is then linear in the averaged square of these fluctuations. The latter in turn are linear in the relevant temperature. Accordingly, the spin current is proportional to the temperature bias.
In the
high-temperature limit (or equivalently a small magnetic field), the magnetic fluctuations are relatively large. Taking into account the asymptotic limit of the Langevin function, $L\big(x\ll1\big)\approx x/3$, in the high magnon temperature limit, the spin current is $\langle I_{tot}^{y}\rangle=(2/3)\alpha' M_{s}V(H_{0}+H_{R})(T_{F}^{m}-T_{N})/T_{F}^{m}$. Thus, the spin current is reduced by the factor $(H_{0}+H_{R})/T_{F}^{m}$, which decreases with increasing magnon temperature or decreasing magnetic field.  Note, the spin current is enhanced when the Rashba and the external fields are parallel
and is reduced in the antiparallel case. Remarkably,  the saturation of the spin current is observed in the high magnon
temperature limit, $T_{F}^{m}\gg T_{N}$, where $\langle I_{tot}^{y}\rangle\approx (2/3)\alpha' M_{s}V(H_{0}+H_{R})$.

Let us assume now a sizable spin-Hall field that cannot be neglected.
The first specific case is when $H_{SH}\approx \alpha\big(H_{0}+H_{R}\big)$.
Taking into account the asymptotic limit of the Langevin function, $L\big(x\ll1\big)\approx x/3$ in the high magnon temperature limit,
$M_{s}V\big(\alpha H_{0} +\alpha H_{R}-H_{SH}\big)/\alpha k_{B}T_{F}^{m}\ll1$, one finds  the following expression for the
 spin current:
\begin{eqnarray}\label{totalspincurrentyequaltemperature0}
&&\langle I_{tot}^{y}\rangle=\frac{2}{3}\alpha' M_{s}V\bigg[\big(H_{0}+H_{R}+\alpha H_{SH}\big)\nonumber\\
&&-\frac{T_{N}}{T_{F}^{m}}\frac{\big(\alpha H_{0}+\alpha H_{R}-H_{SH}\big)}{\alpha}\bigg].
\end{eqnarray}
%Since $\alpha\big(H_{0}+H_{R}\big)\approx H_{SH}$.
Since $\alpha\big(H_{0}+H_{R}\big)\approx H_{SH}$, the second term for any finite $T_{N}/T_{F}^{m}$ in Eq.(\ref{totalspincurrentyequaltemperature0}) is small and can be neglected. Thus, the saturated spin current is
\begin{eqnarray}\label{totalspincurrentyequaltemperature}
&&\langle I_{tot}^{y}\rangle=\frac{2}{3}\alpha' M_{s}V(H_{0}+H_{R}+\alpha H_{SH}).
\end{eqnarray}
The expression for the saturated spin current, Eq.(\ref{totalspincurrentyequaltemperature}), does not depend on  the temperature.
However, Eq.(\ref{totalspincurrentyequaltemperature}) is valid only if  the magnon temperature $T_{F}^{m}$ is high enough.
Thus, by tuning the applied external magnetic field $H_{0}$, a nonzero spin pumping current can be achieved at arbitrary and even at equal temperatures
$T_{F}^{m}=T_{N}$.
For the opposite external  and  Rashba fields, $H_{0}=-H_{R}$, from Eq.(\ref{totalspincurrentyExplicit}) follows
\begin{eqnarray}\label{totalspincurrentopposit}
&&\langle I_{tot}^{y}\rangle=2\alpha' k_{B}L\bigg(\frac{M_{s}VH_{SH}}{\alpha k_{B}T_{F}^{m}}\bigg)\big(\alpha^{2}T_{F}^{m}+T_{N}\big).
\end{eqnarray}
The obtained result is remarkable  as it shows that the net pumping current is finite at arbitrary nonzero temperatures $T_{F}^{m}$ and $T_{N}$ in the absence of the applied temperature gradient.

Finally we explore the case when the fields are comparable $H_{SH}\approx H_{R} \approx H_{0}$, and since $\alpha<1$,  $H_{SH}\gg \alpha H_{R},~\alpha H_{R}$.
\begin{eqnarray}\label{totalspincurrentequal fields}
&&\langle I_{tot}^{y}\rangle=2\alpha' k_{B}L\bigg(\frac{M_{s}V(\alpha H_0 +\alpha H_R -H_{SH})}{\alpha k_{B}T_{F}^{m}}\bigg) \nonumber\\
&&\times \bigg\{\frac{\alpha\big(H_{0}+H_{R}+\alpha H_{SH})T_{F}^{m}}{\alpha H_0+\alpha H_R - H_{SH})}-T_{N}\bigg\}.\nonumber
\end{eqnarray}
The expression of the total spin current in this case reads
\begin{eqnarray}\label{totalspincurrentequal fields}
&&\langle I_{tot}^{y}\rangle=2\alpha' k_{B}L\bigg(\frac{M_{s}VH_{SH}}{\alpha k_{B}T_{F}^{m}}\bigg) \nonumber\\
&&\times\bigg\{\frac{\alpha \big(H_{0}+H_{R}\big)T_{F}^{m}}{H_{SH}}+T_{N}\bigg\}.
\end{eqnarray}
In the low magnon temperature limit we deduce
\begin{eqnarray}\label{totalspincurrentequal fieldsL}
&&\langle I_{tot}^{y}\rangle=2\alpha' k_{B}\bigg\{\frac{\alpha \big(H_{0}+H_{R}\big)T_{F}^{m}}{H_{SH}}+T_{N}\bigg\},
\end{eqnarray}
while in the high magnon temperature limit one finds
\begin{eqnarray}\label{totalspincurrentequal fieldsH}
&&\langle I_{tot}^{y}\rangle=\frac{2}{3}M_{s}V\bigg\{\big(H_{0}+H_{R}\big)+\frac{H_{SH}}{\alpha}\frac{T_{N}}{T^{F}_{m}}\bigg\}.
\end{eqnarray}
As we see from Eq.(\ref{totalspincurrentequal fieldsL}),(\ref{totalspincurrentequal fieldsH}) the role of the field $H_{SH}$ is different.
In the low magnon temperature limit it reduces the spin pumping current, while in the high magnon temperature limit it enhances the fluctuating spin current.

In the analytical calculation, we assumed that temperatures of the magnon subsystem and normal metal are fixed during the process.  However, this is an approximation because the temperatures of the subsystems change slightly during the equilibration process. For illustration, we consider the case when the external and Rashba fields, $H_{0}$ and $H_{R}$,  are parallel and we neglect the spin Hall term. Then from
  Eq.(\ref{totalspincurrentyExplicit}) we deduce:

\begin{eqnarray}\label{equaltemperaturecurrent}
\langle I_{tot}^{y}\rangle=2\alpha'k_{B}L\bigg(\frac{M_{s}V\big(H_{0}+H_{R}\big)}{k_{B}T_{F}^{m}}\bigg)\big(T_{F}^{m}-T_{N}\big).
 \end{eqnarray}
Apparently the total spin current is zero when $T_{F}^{m}=T_{N}$. However, the magnon temperature $T_{F}^{m}$ that we used for derivation of the Fokker-Planck equation, is the initial magnon temperature. The electric current due to the Rashba field modifies the magnon density and magnon temperature, leading to a slight difference in effective magnon temperatures $T_{F}^{m}\big(j_{a}\big)-T_{F}^{m}\big(j_{a}=0\big)=\delta T_{F}^{m}$.
This correction is beyond the Fokker-Planck equation. Therefore, due to the temperature correction $\delta T_{F}^{m}$, in the numerical calculations, we expect to obtain a finite net current even when the initial magnon temperature is
equal to the temperature of the normal metal, $T_{F}^{m}\big(j_{a}=0\big)=T_{N}$.

\section{Spin Current in the N/F/N structure}

\begin{center}
   \begin{figure}[!t]
    \centering
    \includegraphics[width=0.48\textwidth]{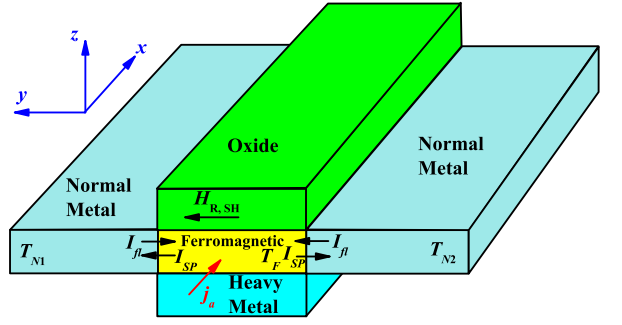}
        \caption{Schematic illustration of the  $N/F/N$ System. The ferromagnetic film is attached to two nonmagnetic layers, $N_{1}$ on the left and
        $N_{2}$ on the right side. The temperatures of the layers $N_1$ and $N_2$ are different. Other notation as in Fig.1.}
    \label{fig2}
    \end{figure}
\end{center}

The same method has been utilized to calculate the spin current in the N/F/N system shown schematically in Fig.{\ref{fig2}}. We calculate the spin current defined as the difference of spin currents flowing through the two interfaces,
\begin{eqnarray}\label{totalSpinCurrentNFN}
&&\vec{I}_{tot}=\vec{I}_{tot1}-\vec{I}_{tot2}\\
&&\vec{I}_{tot1}=\vec{I}_{fl1}+\vec{I}_{sp1}, \vec{I}_{tot2}=\vec{I}_{fl2}+\vec{I}_{sp2}.\nonumber
\end{eqnarray}
 Here $\vec{I}_{tot1}$ and $\vec{I}_{tot2}$ is the total spin current in the first and second interfaces. The total spin current includes four terms. Two terms $\vec{I}_{sp1}$ and $\vec{I}_{sp2}$ describe the spin pumping currents from the ferromagnetic layer to the left $N_{1}$ and to the right $N_{2}$ metallic layers, respectively. In turn, the terms $\vec{I}_{fl1}$ and  $\vec{I}_{fl2}$ describe the fluctuating spin currents flowing from the left and  right metallic layers towards the ferromagnetic layer. We assume that the two metals have different temperatures $T_{N_{1}}$ and $T_{N_{2}}$.
The spin pumping current flowing from the ferromagnetic layer towards metallic layers ($i=1,2$) reads
 \begin{eqnarray}\label{spinpumpcomponents}
 \langle I_{sp1}^{y}\rangle=\alpha'(T_{N_{1}})\frac{M_{s}V}{\gamma}\frac{2\omega_1}{\beta\omega_2}L(\beta\omega_2),\\
 \langle I_{sp2}^{y}\rangle=-\alpha'(T_{N_{2}})\frac{M_{s}V}{\gamma}\frac{2\omega_1}{\beta\omega_2}L(\beta\omega_2).\nonumber
 \end{eqnarray}
 In turn, the fluctuating currents have the components
 \begin{eqnarray}\label{fluctuatingcurrentcomponent}
 \langle I_{fl1}^{y}\rangle=2\alpha'({T_{N_{1}}})k_{B}T_{N_{1}}L(\beta\omega_2),\\
 \langle I_{fl2}^{y}\rangle=-2\alpha'({T_{N_{2}}})k_{B}T_{N_{2}}L(\beta\omega_2).\nonumber
 \end{eqnarray}
 As  we can see from Eq.(\ref{spinpumpcomponents}) and Eq.(\ref{fluctuatingcurrentcomponent}), the difference in the two components of the spin pumping current and fluctuating current is related to the temperature dependence of the damping constant $\alpha'(T_{N})$. For convenience we denote $\alpha'(T_{N_{1}})=\alpha'$ and $\alpha'(T_{N_{2}})=\alpha'+\Delta\alpha$. If the difference between the temperatures of the metals $T_{N_{1}}$ and $T_{N_{2}}$ is not too large, the variation of the damping constant $\Delta\alpha'$ is very small $|\Delta\alpha'|/\alpha'<<1$ \cite{Joyeux,2Chotorlishvili}. In such a case
 \begin{eqnarray}\label{totalspin1totalspin2}
 \langle I^{y}_{tot1}\rangle=2\alpha'k_{B}L(\beta\omega_{2})\bigg(\alpha\frac{\omega_1}{\omega_2}T_{F}^{m}+T_{N_1}\bigg),\\
  \langle I^{y}_{tot2}\rangle=-2\alpha'k_{B}L(\beta\omega_{2})\bigg(\alpha\frac{\omega_1}{\omega_2}T_{F}^{m}+T_{N_2}\bigg),\nonumber
 \end{eqnarray}
 and total spin current:
 \begin{eqnarray}\label{totalspincurrent}
 \langle I^{y}_{tot}\rangle=2\alpha'k_{B}L(\beta\omega_{2})\bigg(2\alpha\frac{\omega_1}{\omega_2}T_{F}^{m}+T_{N_1}+T_{N_2}\bigg).
 \end{eqnarray}
 When $\vec{H}_{\rm eff}=(0,H_0,0)$,  then using Eq.(\ref{correlation1}), Eq.(\ref{correlation2}) and Eq.(\ref{Omega}) one obtains from Eq.(\ref{totalspincurrent}):
 \begin{eqnarray}\label{totalspincurrentyExplicitNFN}
&&\langle I_{tot}^{y}\rangle=2\alpha' k_{B}L\bigg(\frac{M_{s}V(\alpha H_{0}+(\alpha-\eta\xi) H_R-H_{SH})}{\alpha k_B T_{F}^{m}}\bigg) \nonumber\\
&&\times\bigg(2\frac{\alpha(H_{0}+H_{R}+\alpha H_{SH})T_{F}^{m}}{\alpha H_{0}+(\alpha -\eta\xi) H_{R}-H_{SH}}-T_{N_1}-T_{N_2}\bigg).
\end{eqnarray}
When $\eta=0$ and $H_{SH}\ll\alpha H_{0}, \alpha H_{R}$ from Eq.(\ref{totalspincurrentyExplicitNFN}) we get:
\begin{eqnarray}
&&\langle I_{tot}^{y}\rangle=2\alpha' k_{B}L\bigg(\frac{M_{s}V(H_{0}+H_{R})}{k_{B}T_{F}^{m}}\bigg)\times\\
&& \big(2T_{F}^{m}-T_{N_1}-T_{N_2}\big).\nonumber
\end{eqnarray}
Again we see that the larger is the difference between magnon and metal temperatures, the larger is the total spin current.

\section{Effect of DM interaction}

In order to explore the role of DM interaction, we performed micromagnetic simulations for a finite-size N/F system.
To be more specific, we study  Pt/Co/AlO where the Co layer is 500 nm$\times$50 nm large with a thickness of  10 nm. The Co layer is sandwiched between Pt and AlO films.
The parameters describe the Co layer: the saturation magnetization of $M_{s}=10^{6}$A/m, and the damping constant $\alpha=0.2$.
The Rashba field, $H_{R}=\frac{\alpha_{R}P}{\mu_{B}\mu_{0}M_{s}}j_{a}$, can be estimated assuming
$P=0.5$, $\alpha_{R}=10^{-10}$eVm, and the  spatially uniform current density $j_{a}$ along the $x$ axis of the order of $10^{12}$A/m$^{2}$.
Due to the structure of the Rashba field, an increase in the magnitude of current density $j_{a}$ is formally equivalent to the corresponding increase in the SO constant $\alpha_{R}$.
Thus, the dependence of the total spin current on the current density $j_{a}$ is equivalent to the dependence of the total spin current on the SO constant $\alpha_{R}$.

\begin{figure}
     \includegraphics[width=0.48\textwidth]{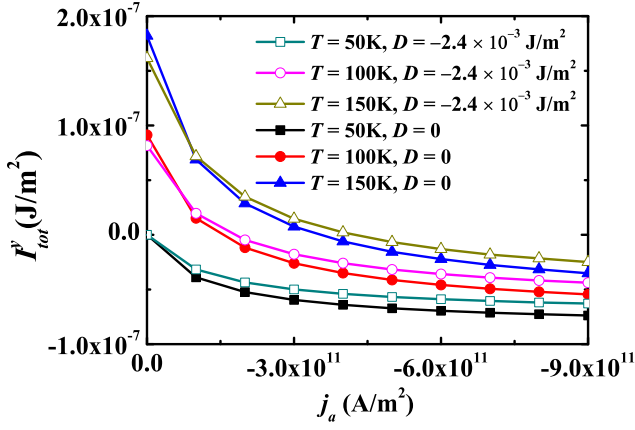}
     \caption{\label{anti-parallel} Total spin current $ I^{y}_{tot} $ in the absence of  the spin Hall effect ($ \theta_{SH} = 0 $), plotted  as a function of the electric current density $ j_{a} $.
 The external field $H_{0}$ and the Rashba field $H_{R}$ are parallel. The magnon temperature  is $ T \equiv T_F^{m} = $ 50 K (squares line), 100 K (circles line) and 150 K (triangles line).
 The DMI constant is assumed $ D = $ 0 (solid dots) and $ D=-2.4 \times 10^{-3} $ J/m$ ^{2} $ (open dots). The external magnetic field $ H_0 = 4 \times 10^5 $ A/m and the normal metal temperature $ T_N = 50 $ K are assumed. }
 \end{figure}

\begin{figure}
     \includegraphics[width=0.48\textwidth]{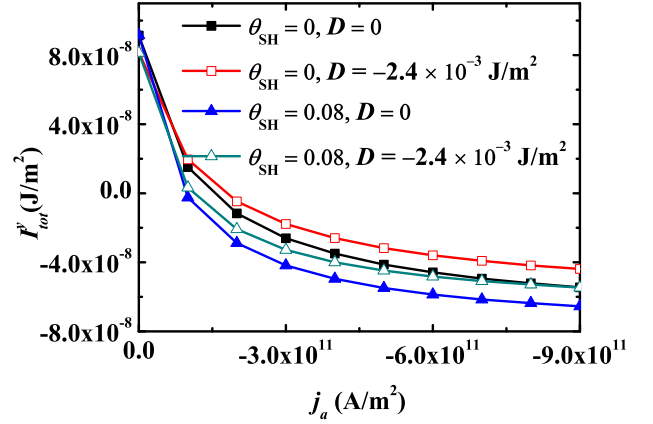}
     \caption{\label{anti-parallel-hall} The total spin current $ I^{y}_{tot} $ without the contribution of the spin Hall effect ($ \theta_{SH} = 0 $, squares line)  and with the spin Hall effect ($ \theta_{SH} = 0.08 $, triangles line), plotted  as a function of the electric current density $ j_{a} $. The external field $H_{0}$ and the Rashba field $H_{R}$ are parallel.  The DM interaction constant is assumed $ D = $ 0 (solid dots) and $ D=-2.4 \times 10^{-3} $ J/m$ ^{2} $ (open dots). The external magnetic field $ H_0 = 4 \times 10^5 $ A/m, the magnon temperature $ T_F^{m} = 100 $ K, and the normal metal temperature $ T_N = 50 $ K are assumed.}
 \end{figure}

In Fig. (\ref{anti-parallel}), the total spin current is plotted as a function of the electric current density $j_{a}$ (assumed negative), for the case when the external magnetic field $H_{0}$ and the Rashba field $H_{R}$ are parallel.
When $j_{a}=0$, the total spin current is solely the spin Seebeck current and is absent for  equal temperatures, $T_F^{m}\big(j_{a}=0\big)=T_N $.
However, when $|j_{a}|> 0$,
the total spin current is nonzero, as well. As it was already mentioned above, the reason of a nonzero net spin current
is a slight shift of the magnon temperature, $T_{F}^{m}\big(|j_{a}|>0\big)-T_{F}^{m}\big(j_{a}=0\big)=\delta T_{F}^{m}$ and of the magnon density, that occur due to the charge current $j_{a}$.
Apparently, in case of antiparallel  Rashba $H_{R}$ and external  $H_{0}$ magnetic fields, the total net spin current decreases with increasing magnitude of the charge current.
This numerical result is consistent with the analytical results obtained in the previous section.
As we see, the effect of the DM interaction is diverse: when $\delta T_{F}^{m}>0$ and spin current is positive $\big<I_{tot}^{y}\big>=\big<I_{sp}^{y}\big>+\big<I_{fl}^{y}\big>>0,~~\big<I_{fl}^{y}\big><0$ (i.e. ferromagnetic layer is {\it hotter} than the normal metal layer), the DM interaction enhances the current.
However, in the case $\delta T_{F}^{m}<0$, when fluctuating spin current is larger than the spin pumping current, and the total net current is negative $\big<I_{tot}^{y}\big><0$, the
DM interaction reduces the spin current. This means that the Rashba $H_{R}$ field always has a positive contribution to the spin pumping current. The situation is the same when the spin Hall effect is included, see Fig. (\ref{anti-parallel-hall}). As one can see, the spin Hall effect has the opposite effect, it always decreases the spin pumping current. Therefore for $\delta T_{F}^{m}>0$ the total spin current without the spin Hall effect is larger, while
for $\delta T_{F}^{m}<0$ it is smaller.

\begin{figure}
     \includegraphics[width=0.48\textwidth]{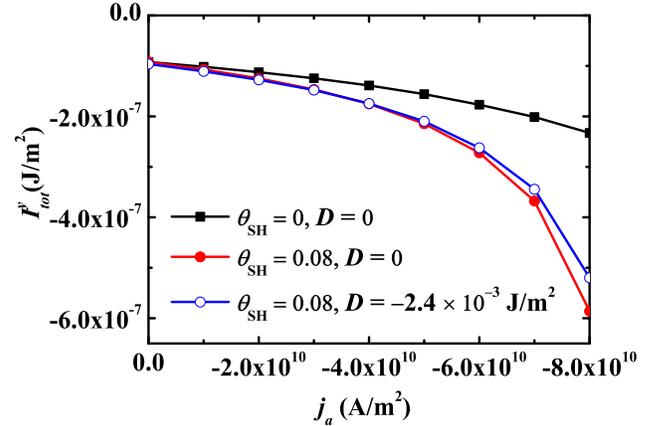}
     \caption{\label{parallel} The total spin current $ I^{y}_{tot} $ with the spin Hall effect ($ \theta_{SH} = 0.08 $, circles line)  and without spin Hall effect ($ \theta_{SH} = 0 $, squares line), plotted as a function of the electric current density $ j_{a} $. The external field $H_{0}$ and the Rashba field $H_{R}$ are antiparallel.  The DMI constant is assumed $ D = $ 0 (solid dots) and $ D=-2.4 \times 10^{-3} $ J/m$ ^{2} $ (open dots). The external magnetic field is $ H_0 = -9 \times 10^5 $ A/m, the  magnon temperature  is $ T_F = 100 $ K, and the temperature of normal metal is $ T_N = 50 $ K. }
 \end{figure}

Finally, we consider the case when the Rashba field $H_{R}$ and the external magnetic $H_{0}$ field are parallel, see Fig. (\ref{anti-parallel-hall}).
Note that a switching of the direction of the magnetic field
alters the ground state magnetic order. Therefore, the spin current changes sign.
As we see from  Fig. (\ref{parallel}), the spin current increases with the electric current density $|j_{a}|$.
This result is also consistent with the analytical result obtained in the previous section.

%\section{Effect of the magnetocrystalline anisotropy}

\begin{figure}
     \includegraphics[width=0.48\textwidth]{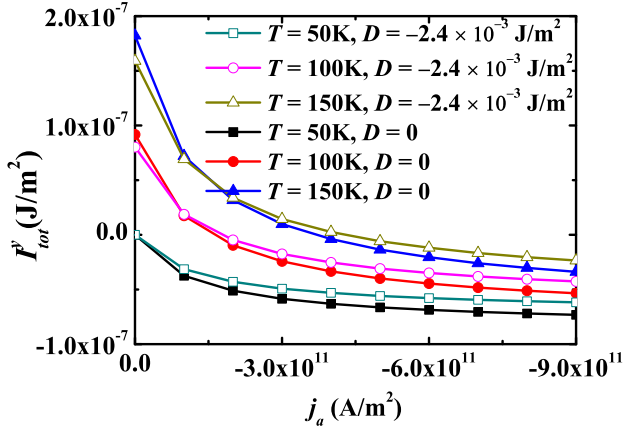}
     \caption{\label{Figure-6} Total spin current $ I^{y}_{tot} $ in the absence of spin Hall contribution ($ \theta_{SH} = 0 $), plotted as a function of the electric current density $ j_{a} $. The local magnetization and the Rashba field are parallel. The magnon temperature is $ T\equiv T_F^{m} = $ 50 K (squares line), 100 K (circles line) and 150 K (triangles line). The DMI constant $ D = $ 0 (solid dots) and $ D=-2.4 \times 10^{-3} $ J/m$ ^{2} $ (open dots). The magnetocrystalline anisotropy constant $ K_y=3 \times 10^{5} $ J/m$ ^{3} $ and the normal metal temperature $ T_N = 50 $ K. The effective anisotropy field is $ \vec{H}_{ani} = 2 K_y m_y \vec{e}_y /(\mu_0 M_s)$. }
 \end{figure}

\begin{figure}
     \includegraphics[width=0.48\textwidth]{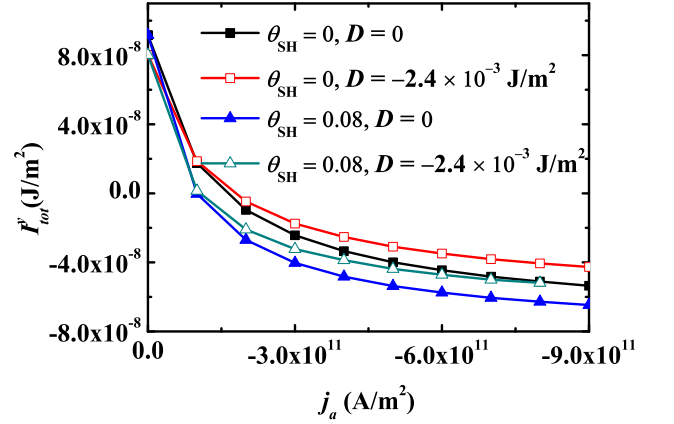}
     \caption{\label{Figure-7} Total spin current $ I^{y}_{tot} $ in the absence of spin Hall effect ($ \theta_{SH} = 0 $, squares line)  and with the Hall effect($ \theta_{SH} = 0.08 $, triangles line), plotted  as a function of the electric current density $ j_{a} $, for the case when the local magnetization and the Rashba field are parallel. The DM interaction constant $ D = $ 0 (solid dots) and $ D=-2.4 \times 10^{-3} $ J/m$ ^{2} $ (open dots). The magnetocrystalline anisotropy constant $ K_y=3 \times 10^{5} $ J/m$ ^{3} $, the magnon temperature $ T_F^{m} = 100 $ K and the normal metal temperature $ T_N = 50 $ K. The effective anisotropy field is $ \vec{H}_{ani} = 2 K_y m_y \vec{e}_y /(\mu_0 M_s)$. }
 \end{figure}

\begin{figure}
     \includegraphics[width=0.48\textwidth]{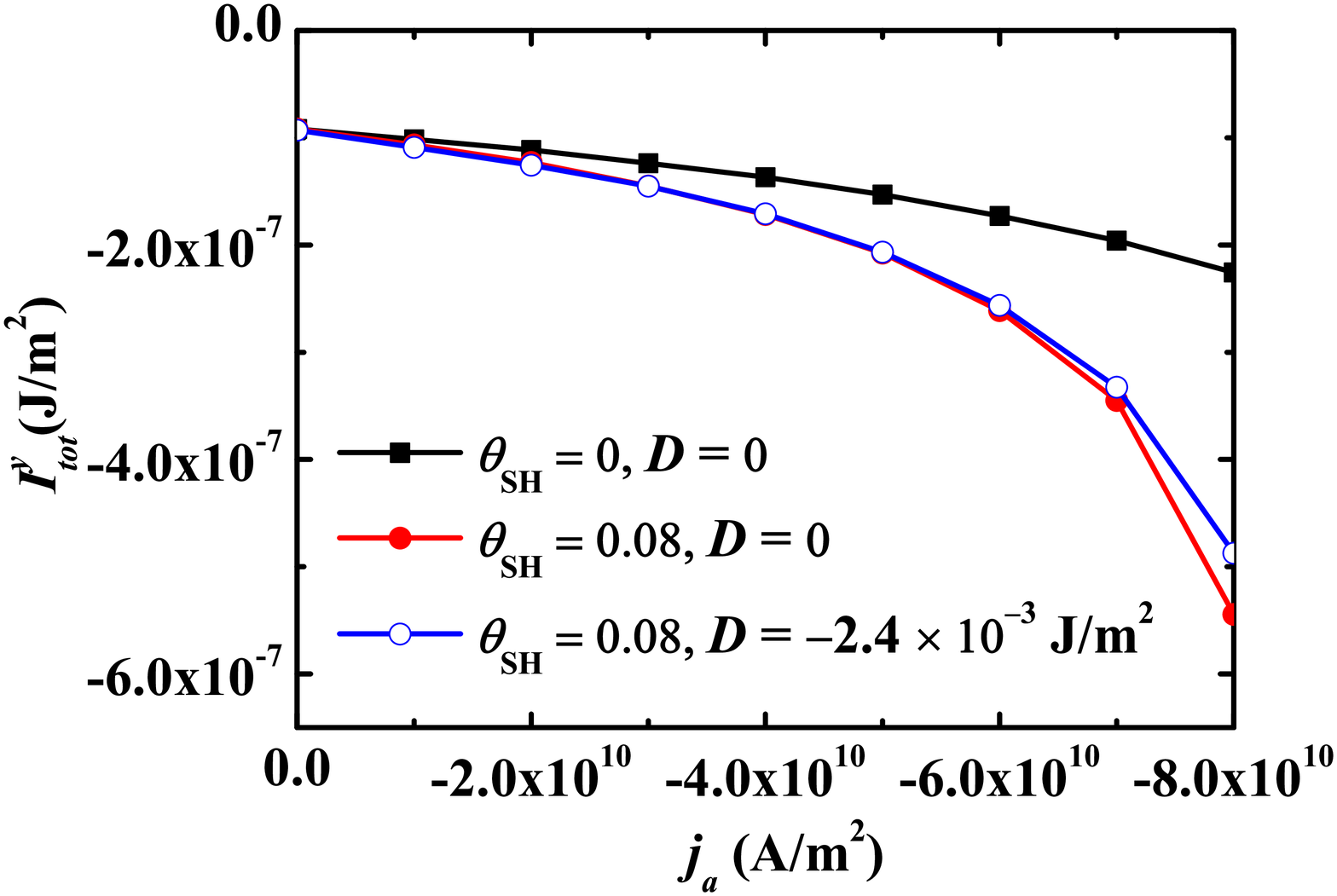}
     \caption{\label{Figure-8} The total spin current $ I^{y}_{tot} $ with the spin Hall effect ($ \theta_{SH} = 0.08 $, circles line)  and without spin Hall effect ($ \theta_{SH} = 0 $, squares line), plotted as a function of the electric current density $ j_{a} $. The local magnetization and the Rashba field are antiparallel.  DMI constant $ D = $ 0 (solid dots) and $ D=-2.4 \times 10^{-3} $ J/m$ ^{2} $ (open dots). The external magnetic field $ H_0 = -5 \times 10^5 $ A/m, the magnetocrystalline anisotropy constant $ K_y=3 \times 10^{5} $ J/m$ ^{3} $, the magnon temperature $ T_F^m = 100 $ K, and the normal metal temperature $ T_N = 50 $ K. The effective anisotropy field is $ \vec{H}_{ani} = 2 K_y m_y \vec{e}_y /(\mu_0 M_s)$. }
 \end{figure}

In order to see the effect of magnetocrystalline anisotropy, we repeated the calculations with the anisotropy term being included. Results of the calculations, plotted in Fig.(\ref{Figure-6}),  Fig. (\ref{Figure-7}), and Fig. (\ref{Figure-8}) show that the magnetocrystalline anisotropy has no significant influence on the spin current, so the effects discussed above hold in the presence of the anisotropy, as well.

\section{Conclusions}
In this paper, we have considered two different heterostructures consisting of a thin ferromagnetic film sandwiched between heavy-metal and oxide layers.
Interfacing the ferromagnetic layer to the heavy metal may result in spin Hall torque exerted on the magnetic moment, while at the interface of the oxide material a spin-orbit coupling of Rashba type emerges. Both factors (the spin Hall effect and the Rashba spin-orbit coupling) have a significant influence on the magnetic dynamics,
and thus also on the spin pumping current. The total spin current crossing the ferromagnetic/normal-metal interface has two contributions:  the spin current pumped from the ferromagnetic metal to the normal one, and the spin fluctuating current flowing in the opposite direction. The spin Hall effect and the Rashba spin-orbit coupling influence only spin pumping current and therefore impact also the total spin current.
We  explored the spin Seebeck current beyond the linear response regime, and found the following interesting features: if the external magnetic field  $H_{0}$ is parallel to the Rashba SO field $H_{R}$, then the SO coupling enhances the spin current, in the case of an antiparallel magnetic field $H_{0}$ and a Rashba SO field $H_{R}$, the
SO coupling decreases the spin current. The spin Hall effect and the DM interaction always increase the spin pumping current. The results are confirmed analytically by means of the Fokker-Planck equation and by direct micromagnetic numerical calculations for a specific sample.

\acknowledgements
This work is supported by the DFG through the SFB 762 and SFB-TRR 227 and
by the National Research Center in Poland as a research project No. DEC-2017/27/B/ST3/02881.

\appendix
\section{Derivation of the Fokker-Plank equation}
For the derivation of the Fokker-Plank equation, we follow Ref.{\cite{Garanin}} and use the functional integration method in order to average the dynamics over all possible realizations of the random noise field.
First we rewrite LLG equation (Eq.(\ref{LLG})) in the form:
\begin{equation}\label{LLG1}
\frac{d\vec{m}}{dt}=-\vec{m}\times(\omega_{1}+\zeta(t))+\vec{m}\times\vec{m}\times\vec{\omega}_{2},
 \end{equation}
 where
 \begin{eqnarray}\label{Omega}
 &&\vec{\omega}_{1}=\vec{\omega}_{eff}+\vec{\omega}_{R}+\alpha\vec{\omega}_{SH},\nonumber \\
 &&\vec{\omega}_{2}=-\alpha\vec{\omega}_{eff}-\alpha\vec{\omega}_{R}+\eta \xi\vec{\omega}_{R}+\vec{\omega}_{SH},\\
 &&\vec{\omega}_{eff}=\gamma\vec{H}_{eff}, \vec{\omega}_R=\gamma\vec{H}_{R},\vec{\omega}_{SH}=\gamma\vec{H}_{SH},\nonumber
 \end{eqnarray}
 and $\gamma\rightarrow \gamma/(1+\alpha^2)$. Here $\vec{\zeta}(t)$ is a random Langevin field with the following correlation properties:
 \begin{equation}\label{Langeven1}
 \langle\vec{\zeta(t)}\rangle=0,
 \end{equation}
 \begin{equation}\label{Langeven2}
 \langle\zeta_{i}(t)\zeta_{j}(t')\rangle=\sigma^{2}\delta_{ij}\delta(t-t').
 \end{equation}
 We introduce the probability distribution function of the random Gaussian noise $\vec{\zeta}$:
 \begin{equation}\label{Gaussian}
 F[\vec{\zeta(t)}]=\frac{1}{Z_{\zeta}}\exp\bigg[-\frac{1}{\sigma^2}\int_{-\infty}^{+\infty}d\tau\zeta^2(\tau)\bigg],
 \end{equation}
 where $Z_{\zeta}=\int D\vec{\zeta}F$ is the noise partition function.
 With the help of Eq.(\ref{Gaussian}) the average of any noise functional $A_{\vec{\zeta}}$ can be written as
 \begin{equation}\label{AverageGaussian}
 \langle A[\vec{\zeta}]\rangle_{\zeta}=\int D\vec{\zeta} A[\vec{\zeta}]F[\vec{\zeta}].
 \end{equation}
Considering the  obvious identity:
 \begin{equation}\label{Identity}
 \frac{\delta\zeta_{\alpha}\tau}{\delta\zeta_{\beta}(t)}=\delta_{\alpha\beta}\delta(\tau-t),
 \end{equation}
 we can calculate first and second variations of $F[\vec{\zeta}(t)]$:
 \begin{equation}\label{firstderivation}
 \frac{\delta F[\vec{\zeta}]}{\delta\zeta_{\alpha}(t)}=-\frac{1}{\sigma^2}\zeta_{\alpha}(t)F[\zeta],
 \end{equation}
 \begin{equation}\label{Secondderivation}
 \frac{\delta^{2}F[\vec{\zeta}]}{\delta\zeta_{\alpha}(t)\delta\zeta_{\beta}(t')}=[\frac{1}{\sigma^4}\zeta_{\alpha}(t)\zeta_{\beta}(t')-\frac{1}{\sigma^2}\delta_{\alpha\beta}\delta(t-t')]F[\vec{\zeta}].
 \end{equation}
 For arbitrary $n$ we have:
 \begin{equation}\label{nderivative}
 \int D \vec{\zeta}\frac{\delta^nF[\vec{\zeta}]}{\delta\zeta_{\alpha_1}(t_1)\delta\zeta_{\alpha_2}(t_2)...\delta\zeta_{\alpha_n}(t_n)}=0.
 \end{equation}
Taking into account  Eq.(\ref{firstderivation}) to Eq.(\ref{nderivative}), we obtain (\ref{Langeven1}) and (\ref{Langeven2}).
Now, we introduce the  distribution function:
\begin{equation}\label{distribution}
f(\vec{N},t)=\langle\vec{\pi}([\vec{\zeta}],t)\rangle_{\zeta}, ~~~\vec{\pi}([\vec{\zeta}],t)=\delta(\vec{N}-\vec{m}(t)),
\end{equation}
on the sphere $|\vec{N}|=1$.
Taking into account the relation \cite{Garanin} $\dot{\vec{\pi}}=-\frac{\partial\vec{\pi}}{\partial\vec{N}}\dot{\vec{m}}(t)$
and the equation of motion, Eq.(\ref{LLG1}), we deduce  the Fokker-Plank equation:
\begin{eqnarray}\label{Fokker-plank}
\frac{\partial f}{\partial t}=&&\frac{\partial}{\partial\vec{N}}[(\vec{N}\times \vec{\omega}_{1})-(\vec{N}\times\vec{N}\times\vec{\omega}_{2})\nonumber\\
&&+\vec{N}\times\langle\vec{\zeta}(t)\vec{\pi}([\vec{\zeta}],t)\rangle_{\zeta}].
\end{eqnarray}
To calculate $\langle\vec{\zeta}(t)\vec{\pi}([\vec{\zeta}],t)\rangle_{\zeta}$ we use the standard procedure, discussed for example in Ref. \cite{Garanin}, and obtain
\begin{equation}\label{procedure}
\langle\vec{\zeta}(t)\vec{\pi}([\vec{\zeta}],t)\rangle_{\zeta}=-\frac{\sigma^2}{2}\vec{N}\times\frac{\partial f}{\partial \vec{N}}.
\end{equation}
The Fokker-Plank equation in the final form reads
\begin{eqnarray}\label{Fokker-plank equation}
& \frac{\partial f}{\partial t}=\frac{\partial}{\partial\vec{N}}[(\vec{N}\times \vec{\omega}_{1})\nonumber \\
& -(\vec{N}\times\vec{N}\times\vec{\omega}_{2})-\frac{\sigma^2}{2}\vec{N}\times\frac{\partial f}{\partial \vec{N}}].
\end{eqnarray}
The stationary solution of the Fokker-Plank equation when $\vec{\omega_{1}}||\vec{\omega_{2}}$ has the form
\begin{equation} \label{Stationarysolution}
f(\vec{N})=\frac{e^{-\frac{2}{\sigma^2}\int d\vec{N}\cdot\vec{\omega_2}}}{\int d\vec{N} e^{-\frac{2}{\sigma^2}\int d\vec{N}\cdot\vec{\omega_2}}}.
\end{equation}

 \section{Mean values of magnetization}

Exploiting  the parametrization:
 \begin{eqnarray}\label{parametrization}
 &&m_{x}=\sin\theta\cos\varphi,m_{y}=\sin\theta\cos\varphi,m_{z}=\cos\theta ,\nonumber \\
&& 0\leq\theta\leq\pi,0\leq\varphi\leq 2\pi,
 \end{eqnarray}
and taking into account Eq.({\ref{Stationarysolution}}) and the parametrization Eq.(\ref{parametrization}), we can write the probability distribution for $\vec{m}$:
 \begin{eqnarray}\label{Probability Distribution}
 &&dw(\theta,\varphi)=\frac{1}{Z}f(\theta,\varphi)d\vec{m},\\
 &&f(\theta,\varphi)=\exp(-\beta\omega_2\sin\theta\sin\varphi),\nonumber \\
 &&dm=\sin\theta d\theta d\varphi, \beta=\frac{2}{\sigma^2}.\nonumber
 \end{eqnarray}
Here $Z=\frac{4\pi\sin(\beta\omega_2)}{\beta\omega_2}$ is the partition function.
 From Eq.(\ref{Probability Distribution}) we can calculate the mean values of  the magnetization:
 \begin{eqnarray}\label{mean valuea magnetization}
 &&\langle m_x\rangle=\langle m_z\rangle=0,\langle m_y\rangle=-L(\beta\omega_2)\\
 &&\langle m_x^2\rangle=\langle m_z^2\rangle=\frac{L(\beta\omega_2)}{\beta\omega_2}, \langle m_y^2\rangle=1-\frac{2L(\beta\omega_2)}{\beta\omega_2}\nonumber \\
 &&\langle m_x m_y\rangle=\langle m_z m_y\rangle=0. \nonumber
 \end{eqnarray}
 \\

 \section{Derivation of Eq.(17)}
To calculate (\ref{nonzerotransform}) we utilize the Jourdan's lemma,
 \begin{eqnarray}\label{Integral}
 &&\int_{-\infty}^{+\infty}\frac{\omega_2\langle m_y\rangle+i\omega}{(\omega_2\langle m_y\rangle+i\omega)^2+\omega_1^2}e^{-i\omega t}\frac{d\omega}{2\pi}\\
 &&=\int_{-\infty}^{+\infty}\frac{\omega_2\langle m_y\rangle-i\omega}{(\omega_2\langle m_y\rangle-i\omega)^2+\omega_1^2}e^{i\omega t}\frac{d\omega}{2\pi}\nonumber\\
  &&= \left\{ \begin{array}{ll}
         -\frac{1}{2}(e^{-i(\omega_1+i\omega_2\langle m_y\rangle)t}+e^{i(\omega_1-i\omega_2\langle m_y\rangle)t}) & \mbox{if $t> 0$},\\
        0 & \mbox{if $t < 0$}.\end{array} \right.\nonumber
 \end{eqnarray}
 This integral is discontinuous at $t=0$, therefore the value $\langle \vec{m}(t)\times \vec{\zeta'(0)}\rangle_y$ at $t=0$ is given by the average of the values at $t=0^{\pm}$. Therefore, from Eq.(\ref{Integral}) we deduce Eq.(\ref{Corelator})
 \begin{eqnarray}
 &&\langle \vec{m}(0)\times \vec{\zeta}'(0)\rangle_y=\sigma'^2\langle m_y\rangle =-\sigma'^2 L(\beta\omega_2) .\nonumber
 \end{eqnarray}
 
\end{document}